\documentclass[conference]{IEEEtran}
\usepackage{amsmath, amsthm, amssymb}
\usepackage{graphicx}
\usepackage{multirow}
\usepackage[ruled,vlined]{algorithm2e}

\begin{document}

\title{Tagging with DHARMA, a \textit{DH}T-based \textit{A}pproach for \textit{R}esource \textit{M}apping through \textit{A}pproximation}

\author{Luca Maria Aiello, Marco Milanesio, Giancarlo Ruffo and Rossano Schifanella\\
Computer Science Department - Universit\`a degli Studi di Torino\\ 
\{aiello, milane, ruffo, schifane\}@di.unito.it\\
}

\maketitle
\thispagestyle{empty}

\begin{abstract}
We introduce collaborative tagging and faceted search on structured P2P systems. Since a trivial and brute force mapping of an entire folksonomy over a DHT-based system may reduce scalability, we propose an approximated graph maintenance approach. Evaluations on real data coming from Last.fm prove that such strategies reduce vocabulary noise (i.e., representation's overfitting phenomena) and hotspots issues.
\end{abstract}

\section{Introduction}
Social applications are rapidly popularizing collaborative tools for indexing, retrieval, access and distribution of content over the Internet. Multimedia resources are made available through websites and p2p systems, together with annotations, metadata, tags, and other kind of information about the owner and/or the content itself. Such information is often used to fill the semantic gap between the personal user experience, and a more general description of a given resource. Nevertheless, such huge volume of information is often hidden to traditional search engines, since a common query infrastructure and language is missing. 

During the years, the Web community has been supported with many retrieval techniques, that can be categorized in two main paradigms: \textit{navigational search} and \textit{direct search}. The first family of strategies assumes the existence of a taxonomy, usually predefined by a group of experts, that can be iteratively browsed by a user from general categories to more specific subclasses of information (e.g., Yahoo! Directory). Direct search let the user query the engine by means of a (set of) keyword(s) (e.g. Google). Even if the latter has gained a vast amount of success during the last years, very recently navigational paradigm has emerged again due to the diffusion of folksonomies within popular tagging systems (e.g., Flickr, del.icio.us, and so on); in fact, folksonomies have been showed to overperform monolithic hierarchical classifications in social domains where many users with different mental attitudes and vocabularies are active.
Quite surprisingly, in the p2p domain navigational search's benefits have been understimated, and few proposals exist in the related literature. In particular, a lot of effort has been devoted to direct search strategies, very common in unstructured p2p systems, and to \textit{exact match key-based lookup} techniques, that are basically used by almost every structured overlay network. Even if some scholars have proposed semantic routing for p2p systems (moving from the pioneering work of Crespo and Garcia-Molina \cite{crespoG04}), few research has been conducted on merging collaborative tagging, folksonomies and p2p systems. 

First of all, we need to adopt a general tagging system model that can be exploited to define navigational search strategies (Section \ref{model}). Such model should fit the social media domain in the broadest sense of the word, since it could be used to implement a high level engine that allows the user to search in different environments (web, social networks, p2p file sharing networks, and so on). 

A structured p2p system is the natural setting for implementing such model, because of better scalability, and inherent distribution of keys and indexes of resources. Nevertheless, it is hard to find a one-to-one mapping of a given folksonomy (seen as a network of tags) and a DHT system (that partitions a given keyspace among the participating nodes). In Section \ref{implementation} we propose a way to perform such a mapping, introducing an approximation strategy that fits well with dynamic and decentralized tagging.

Finally, a very relevant issue may arise if unbalanced distributions of popular tags are used. As we show in Section \ref{eval}, our proposal can be used efficiently in a real domain (i.e., Last.fm), due to the approximation strategies cited above.

\section{Related works}\label{back}

Several other efforts in studying the possible deploying of collaborative tagging systems on peer to peer environments has been recently made.

In \cite{asiki08support} an efficient indexing scheme for storing and retrieving concept hierarchies over a fully decentralized system is given, even if folksonomies are not taken into account.

A p2p infrastructure for tagging systems (\textit{PINTS}) is proposed in \cite{gorlitz08pints}; in particular, the authors design a scheme to maintain feature vectors for characterization of users and resource of a tagging environment on a DHT. Feature vectors may be useful for calculating the similarity between users  or for constructing algorithms for ranked retrieval.

PINTS comes as a building block of Tagster \cite{gorlitz08tagster}, a distributed content sharing and tagging system where the user-resource-tag graph is stored in a DHT. A dedicated storing index is used for each tagging relation, so each edge in the graph is stored at different overlay responsibility areas. For this reason, one lookup for each edge retrieval is needed, and this could make the navigation expensive in systems with a huge number of tags and objects. Furthermore, navigational aspects between related tags is not explicitly taken into account.

In \cite{mozo08scalable} \textit{T-DHT}, an hybrid structured-unstructured p2p approach is described. The scheme does not explicitly model a folksonomy with inter-tag relations and with the possibility to navigate through related labels.

Tag-based navigation is taken into account in \cite{bouillet08folksonomy-based}, where a centralized web service discovery system based on folksonomies is presented. Tags, together with variables, are used to assign semantic information to input and output messages of the service operations. The key feature of this work is the possibility to exploit the subsumption relation between variable types to compose the discovery activity as an acyclic navigational workflow.

Even if the work described in \cite{heymann06collaborative} is not related to p2p, it worths to be cited here because of the link the authors put between taxonomies and folksonomies. It has inspired us to further insights on query convergence.

\section{Tagging system model}\label{model}

The first step in the definition of our distributed search engine is an high level description of the tagging system; we define the tag-resource graph and also a folksonomy graph by means of a simple tag similarity measure (Section \ref{model:structure}), and how these graphs are modified during users interaction (Section \ref{model:maintenance}); finally, we show how this model can be used for navigating through tags in search of resources (Section \ref{model:navigation}).

\subsection{Graphs definition}\label{model:structure}

Usually, collaborative tagging systems are defined as tripartite hypergraphs \cite{lambiotte06collaborative, mika07ontologies}, in which three sets of actors are involved:
\begin{itemize}
\item $U$ is the set of users of the system, that actually tag resources.
\item $T$ is the set of tags.
\item $R$ is the set of resources being tagged.
\end{itemize}

\begin{figure}[tp]
	\centering
		\includegraphics[scale=.65, keepaspectratio]{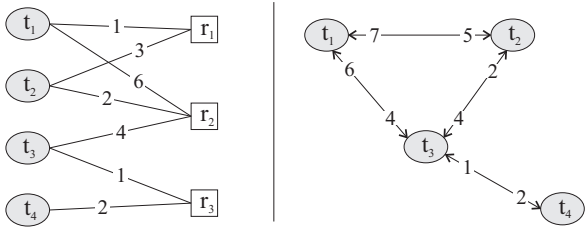}
	\caption{Bidirectional arcs in a Folksonomy Graph (right) aggregates asymmetrically weights in the Tag-Resource graph (left)}
	\label{fig:tag_graph}
\end{figure}
However, since in our work we focus mainly on tags and resources, we perform an aggregation across the user dimension in order to obtain a bipartite graph that links tags to resources. We define such a graph as the \textit{Tag-Resource Graph (TRG)}, where $TRG = (T \cup R, E_{TR})$, s.t. $(t,r) \in E_{TR}$ iff at least a user tagged $r$ with $t$. Moreover, for each arc in $E_{TR}$ we define a weight $u(t,r)$ that is the number of times $r$ has been tagged with $t$ (see Figure \ref{fig:tag_graph} on the left). The reader can observe that we are adopting the so-called \textit{distributional aggregation approach} \cite{markines09similarity} that yields to a graph in which the weight of an edge $(t,r), t \in T, r \in R$ is equal to the number of users tagging $r$ with $t$.

We can use such graph to extract $Tags(r)$ and $Res(t)$ denoting, respectively, the subset of tags that label a resource $r$ and the subset of resources that have been tagged with $t$:
\begin{eqnarray}
	Tags(r) = \{t \in T | \exists (t,r) \in E_{TR} \}, r \in R \label{eq:1}\\
	Res(t) = \{r \in R | \exists (t,r) \in E_{TR} \}, t \in T \label{eq:2}
\end{eqnarray}

Since our purpose is to define a tag-based search engine, we introduce a simple \textit{Folksonomy Graph (FG)} that can be trivially derived through collaborative tagging. Intertag correlations should be detected by means of a distance measure between any pair of tags. We interpret such a distance as an asymmetric similarity function between two generic tags $t_1$ and $t_2$, s.t. $sim(t_1, t_2) = \displaystyle\sum_{r \in Res(t_1)} u(t_2,r)$.\\
Roughly, $sim(t_1, t_2)$ says how many times resources labeled with $t_1$ have been tagged also with $t_2$. Even if many different similarity measures could be adopted in folksonomies \cite{markines09similarity}, such aggregation of tag-resource weights is a metric that is easy to calculate and, as we explain in Section \ref{implementation}, handy to be mapped on a fully decentralized and dynamic context. This metric can be considered as a generalization of tag-tag co-occurrence \cite{cattuto08semantic}. 
\begin{figure}[tp]
	\centering
		\includegraphics[scale=.63, keepaspectratio]{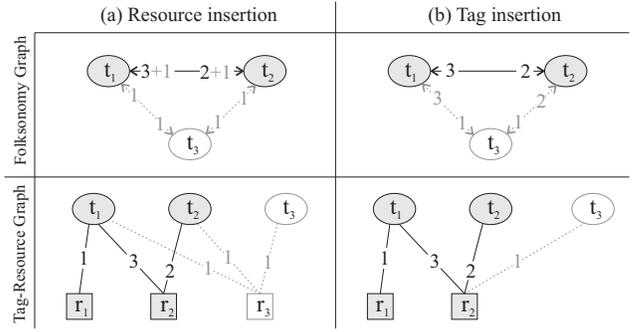}
	\caption{(a) Resource insertion: resource $r_3$, labeled with $t_1$,$t_2$,$t_3$ is inserted (b) Tag insertion: tag $t_3$ is attached to $r_2$. Light arcs and nodes are those added during the insertion}
	\label{fig:graph_maintenance}
\end{figure}

Now, we can define our folksonomy graph as $FG = (T, E_F)$, s.t. $(t_1, t_2) \in E_F$ iff $sim(t_1, t_2) \geq 1$. Let us observe that, by construction, if $sim(t_1, t_2) \neq 0$, then $sim(t_2, t_1) \neq 0$, even if it may happen that $sim(t_1, t_2)\neq sim(t_2, t_1)$. Hence, we represent connections between tags using bidirectional arcs with two weights. Finally, we will need to deal with the tags related to a given tag $t$. Such set is the neighborhood of $t$ in $FG$, denoted with $N_{FG}(t)$. 

For example, in Figure \ref{fig:tag_graph} (right), the arc $(t_1,t_2)$ of FG has weight $5$ because resources $r_1$ and $r_2$, have been tagged also with $t_2$ by, respectively, $3$ and $2$ users; let us observe that, conversely, $sim(t_2,t_1)=7$. 

\subsection{Graphs maintenance}\label{model:maintenance}

The active collaborative behavior of the user community leads to a continuous evolution of the TRG and the FG, due to the addition of new items and new annotations.

\subsubsection{Resource insertion}

When an user inserts a new item $r$ and tags it with $T_r = \{t_1, ... , t_m\}$, then a new resource vertex is inserted in the TRG. Of course, also for each new tag $t_i$, a new vertex is inserted in the TRG, so that $R$ is updated to $R \cup \{r\}$, and the set of tags to $T \cup T_r$. Moreover, for each $t_i \in T_r$, an edge $(r,t_i)$ is added to $E_{TR}$, with $u(r,t_i) = 1$.
As a consequence, $FG$ must be changed, too: for each pair of tags $t_i,t_j \in T_r$, a new arc $(t_i,t_j)$, if not previously existent, is added to $E_F$ with $sim(t_i,t_j)=sim(t_j,t_i)=1$. Otherwise, $sim(t_i,t_j)$ and $sim(t_j,t_i)$ are simply incremented of one unit.

\subsubsection{Tag insertion}
Graphs grow also when an existent resource $r$ is tagged with $t$. Firstly, if $t \notin T$, a proper tag node is added. Therefore, if $t \notin Tags(r)$, then a new edge $(t,r)$, with $u(t,r)=1$ is added to $E_{TR}$; conversely, if $t \in Tags(r)$, then $u(t,r)$ is simply incremented. Similarities are changed consequently. For each tag $\tau \in Tags(r)$, $sim(\tau,t)$ is incremented by one. Instead, $sim(t,\tau)$ is changed depending on whether $t$ was in $Tags(r)$ before the tagging operation or not. If $t$ was in $Tags(r)$, then $sim(t,\tau)$ is left unchanged, otherwise $sim(t,\tau)$ is incremented by $u(\tau,r)$. Arcs in the $FG$ are created or updated accordingly.

Examples of both operations are showed in Figure \ref{fig:graph_maintenance}.

\subsection{Faceted Search within the Folksonomy Graph}\label{model:navigation}

Our purpose is to exploit our model in order to let the user explore the given multi-dimensional information space by iteratively narrowing the number of choices at each search step.

Many popular tagging systems (e.g., Flickr, Last.fm, and so on) make use of resource \textit{clustering} according to some measure of similarity. For example, considering our Folksonomy Graph, we can easily identify clusters representing repeated patterns of tags that can be presented to the users through lists or tag clouds. Such clusters can be intuitively used to refine the query or to disambiguate search keywords. Nevertheless, clustering techniques can generate unpredictable groups, produce cycles in the navigation process, and limit the browsing features of the system since they may not allow refinements.

Generally speaking, users prefer hierarchical classifications with clear and meaningful labels at each level of the tree. For example, a tag that is presented more than once during the same search process can generate confusion, as well as a general term (e.g., ``rock'') that is found in a cluster after that a specific tag (e.g., ``heavy-metal'')  has been selected. Unfortunately, a traditional and rigorous taxonomy is difficult to be provided in a highly dynamic social domain with many users with different mental attitudes and vocabularies.

\textit{Faceted search} can be seen as a middle ground approach that allows the user to ``dive'' the folksonomy without semantic cycles and to iteratively refine the tag-resources space (e.g., TagExplorer by Yahoo! Research). Accordingly to this approach, the user browses the tagging system through a path in FG. We can interpret every tag of such a path as a different level of a hierarchical faceted search process; in fact, selecting subsequent tags in the hierarchy results in a conjunction over the selected annotations, and each step zooms in the tag-resource space, narrowing the focus of the search. 

Two important consequences of this approach are \textit{query convergence} and \textit{vocabulary specialization}. Let us assume that the user starts the search process selecting tag $t_0$, and afterwards she chooses $t_1, t_2, \ldots, t_n$. At each step, only co-related tags are presented to the user, i.e., we have that $t_i$ is always a neighbor of $t_{i-1}$ in FG, that is $t_i \in N_{FG}(t_{i-1})$. Moreover, at step $i$ a set of tags $T_i$ and a set of resources $R_i$ can be presented to the user:
\begin{small}\[  
T_i = \begin{cases}N_{FG}(t_0) & i = 0\\T_{i-1} \cap N_{FG}(t_i) & i > 0\end{cases};
R_i = \begin{cases}Res(t_0) & i = 0\\R_{i-1} \cap Res(t_i) & i > 0\end{cases}
\]\end{small}
Even if we are not concerning on presentation aspects, we can assume that to improve usability only a subset of $T_i$ is displayed (using a tag cloud or an alternative representation), and that a subsequent tag selection would be equivalent as a zoom in, eventually visualizing other (more specific) tags. Obviously, since previously chosen tags are not taken into account in subsequent steps, $\forall i: |T_i| < |T_{i-1}|$. The upper bound of the iterative process is $O(|T_0|)$, and so convergence is trivially proved. It can be noted that browsing can be delayed by tags that are ``semantically equivalent'' (i.e., all $\tau \in T_{i-1}$ s.t. $|T_i| = |T_{i-1}|$). However, such situations are limited in numbers and do not affect significantly search performance.

\section{Mapping on a DHT}\label{implementation}

Next, we present a general insight of how the model defined in Section \ref{model} can be mapped on a Distributed Hash Table. We show that a naive implementation of our model would lead to grievous inefficiencies which severely limit the scalability of the system; therefore, we propose an approximated approach to overcome these issues.

\subsection{Distributed model}\label{implementation:mapping}

In order to map the folksonomy on a DHT we need to shrink the TRG and the FG both in small structural \textit{blocks} that can be stored at different overlay nodes. In particular, each block contains a node together with its outgoing edges. Accordingly, every resource node $r \in TRG$, together with its outgoing edges to nodes $t \in Tags(r)$ is contained into a single block. Symmetrically, every tag $t \in TRG$ with its outgoing edges to $r \in Res(t)$ forms a block. Likewise, the FG is partitioned in blocks containing a tag $t$ with the arcs that links it to its neighbors in $N_{FG}(t)$. More formally, we define four types of blocks:

\begin{enumerate}
	\item $\bar{r}$ : $\{(t, u(t,r)) | t \in Tag(r)\}, r \in R$
	\item $\bar{t}$ : $\{(r, u(t,r)) | r \in Res(t)\}, t \in T$
	\item $\hat{t}$ : $\{(t',sim(t,t')) | t' \in N_{FG}(t)\}, t \in T$
	\item $\tilde{r}$ : $(r,URI(r)), r \in R$
\end{enumerate}

The TRG is split into blocks of type 1 and 2, the FG into blocks of type 3. Type 4 blocks are introduced only to conceptually associate the resource itself (a URI of a generic object or service) to its name $r$ (a human readable identifier which denotes the resource). Each block is mapped on a lookup key computed from the name of its node concatenated with a string which determines the block type (e.g. the hash of $t | $``$2$'' is the key of type 2 block for tag $t$). For brevity, we denote with $\bar{r}$, $\bar{t}$, $\hat{t}$, $\tilde{r}$ the lookup keys for blocks of type 1-4; for simplicity, we use this notation to directly denote the blocks without introducing any ambiguity.

\begin{figure}[tp]
\centering
\includegraphics[scale=.60, keepaspectratio]{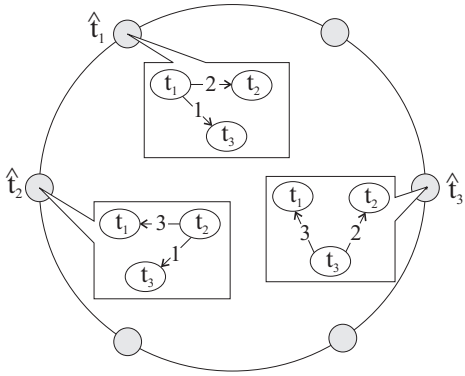}
\caption{Folksonomy graph mapping on the DHT}
\label{fig:dht-mapping-FG}
\end{figure}

\begin{figure}[tp]
\centering
\includegraphics[scale=.60, keepaspectratio]{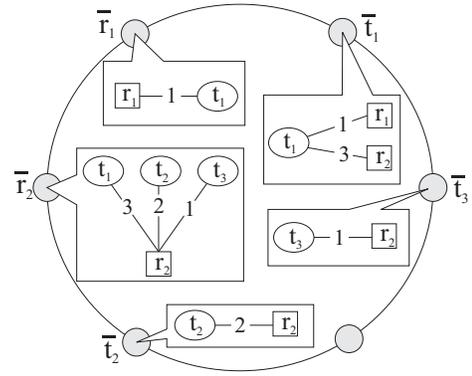}
\caption{Tag Resource graph mapping on the DHT}
\label{fig:dht-mapping-TRG}
\end{figure}

Figures \ref{fig:dht-mapping-FG} and \ref{fig:dht-mapping-TRG} shows how the FG and the TRG depicted in Figure \ref{fig:graph_maintenance}.b are partitioned in blocks and mapped on a generic DHT layer. Overlay nodes are labeled with the key they are responsible for; the content of overlay nodes' storages are depicted into the baloons. Type 4 blocks are omitted for simplicity.

Given such mapping, navigation, tagging and resource insertion through the p2p network are easy to describe. At each navigation step, when a tag $t$ is selected, tags and resources related to $t$ are retrieved by fetching blocks $\hat{t}$ and $\bar{t}$; intersection with tag and resources set retrieved in following steps are performed locally. Insertion of a resource $r$, marked with tags $t_i, i \in {1..m}$, requires the creation of block $\tilde{r}$ to store the URI and of block $\bar{r}$ to connect the resource with its tags. Reverse tag-resource connections are mapped by inserting blocks $\bar{t}_i$ for each tag $t_i$ given in input. Each $t_i$ should then be connected to others in the FG by creating (or updating) its block $\hat{t}_i$. Finally, when a resource $r$ is tagged with a label $t$, the weight of edge $(t,r)$ is incremented by updating blocks $\bar{r}$ and $\bar{t}$. Then, tags $\tau \in Tags(r)$ are retrieved from block $\bar{r}$. For every $\tau$, the weights of arc $(t,\tau)$ is incremented by updating block $\hat{t}$, while reverse connections $(\tau,t)$ must be updated by modifying blocks $\hat{\tau}, \forall \tau \in Tags(r)$.

We suppose that retrieving or modifying the content of a block on the DHT costs only one overlay lookup operation. This assumption is reasonable if the overlay is equipped with proper \textsc{put} and \textsc{get} operations, which, respectively, insert and retrieve contents from the DHT by exploiting the overlay network's lookup service.
In particular, we suppose that a block's structure is modified only by the addition (or, possibly, deletion) of \textit{one-bit tokens}, which determines a unit increment of a particular arc in the TRG (or FG). This approach leads to a more simple implementation, suitable for any DHT; however we omit implementation details due to space limitations. We implemented such primitives on Likir \cite{aiello08tempering}, based on Kademlia \cite{maymounkov02kademlia}. Given this assumptions, we can easily calculate the cost of each basic operation on the folksonomy; results are listed in the first row of Table \ref{tab:complexity}.

\subsection{Approximated approach}\label{implementation:approx}

Implementing the algorithms defined in Section \ref{model:maintenance} with our distributed framework produces two severe issues, both concerning the tagging operation (i.e. the FG update).

The first is a complexity problem. We stated that when a new tag $t$ is added to a resource $r$ the weights of the arcs $(\tau,t), \tau \in Tags(r)$ must be updated. In the DHT domain this implies the update of blocks $\hat{\tau}$ of each $\tau \in Tags(r)$. Accordingly, a number of lookups which is linear with $|Tags(r)|$ is performed. This cost is unsustainable because, as we show later in Section \ref{eval}, a resource can be tagged with several hundred labels. Even if different lookups can be executed in parallel, the bandwidth usage would be definitely excessive for such simple and frequent operations.

The second is a consistency problem, caused by a \textit{race condition}. To keep the graph consistent with our model, if the arc $(t,\tau), \tau \in Tags(r)$ was not present before new tag $t$ insertion, then $sim(t,\tau)$ should be incremented by $u(\tau,r)$. Nevertheless, it is hard to implement correctly this practice in a fully decentralized system. It is easy to understand, indeed, that if two users try to add simultaneously the same tag $t$ on the same resource $r$, there is the risk that the value of $sim(t,\tau)$, for any $\tau \in Tags(r)$, is uncorrectly incremented twice, for a total value of $2 \cdot u(\tau,r)$.

These considerations must be taken into account to improve the algorithm design. We adopted two approximated strategies to solve these problems; call $t$ the new tag and $r$ the resource to be labeled.

\noindent \textbf{\textit{Approximation A}}. Instead of incrementing the weights of all the arcs $(\tau,t), \tau \in Tags(r)$, perform the increment only for a \textit{random subset} of $Tags(r)$. The cardinality of such subset can be chosen to be at most a \textit{constant} number $k$; this expedient reduces the number of lookups needed for a tagging operation, preventing its complexity to scale with $|Tags(r)|$. We refer to $k$ as the \textit{connection parameter} of the approximated graph $\square$

\noindent \textbf{\textit{Approximation B}}. If the arc $(t,\tau), \tau \in Tags(r)$ was not present before the tagging operation, then increment the weight of $(t,\tau)$ only by \textit{one} (and not by $u(\tau,r)$). This avoids the possible inconsistencies due to simultaneous addition of a new tag $t$ to resource $r$ $\square$
\begin{table}[tp]
\centering
	\begin{tabular}{|c|c|c|c|}
		\hline
		\textbf{Primitives} & \textbf{Insert ($r , t_{1..m}$)} & \textbf{Tag ($r$,$t$)} & \textbf{Search step}\\
		\hline
		\#lookups (naive) & $2 + 2m$ & $4 + |Tags(r)|$ & $2$\\
		\hline
		\#lookups (approx.) & $2 + 2m$ & $4 + k$ & $2$\\
		\hline
	\end{tabular}
	\caption{Distributed tagging system primitives cost}
	\label{tab:complexity}
\end{table}

Approximations make the similarity graph evolve differently from the abstract model described in Section \ref{model}. Thus, the distance from the theoretic and the mapped graph should be measured to check how much the search procedure is affected by our approximations. In Section \ref{eval} we present experimental results to measure such distance. It is worth noting that only the FG is affected by the approximation, while the TRG graph remains the same. Complexity of approximated operations in terms of overlay lookups is shown in the second row of Table \ref{tab:complexity}.

The source code of the distributed tagging application that implements the approximated approach is available online\footnote{http://likir.di.unito.it}, with the name of \textit{DHARMA} (\textit{DH}T-based \textit{A}pproach for \textit{R}esource \textit{M}apping through \textit{A}pproximation). An implementation of the underlying DHT is available as well.

\section{Evaluation}\label{eval}

We give an evaluation on how the approximations introduced in our system design impact on the validity of the model using analytic and simulative approaches. The analysis is based on a dataset extracted from Last.fm. First (Section \ref{eval:overview}), we give a brief description of the main features of the dataset, then (Section \ref{eval:experiments}) we analyze how the FG created through the protocol we defined in Section \ref{implementation:approx} well approximates the theoretic similarity model of the dataset and we show that user search experience does not decay due to introduced approximations. Additionally, in Section \ref{eval:convergence} we report the results of a simulative experiment aimed at the estimation of the mean number of steps needed for query convergence.

We base our experimental analysis on a snapshot of the Last.fm web site collected from January 2009 to April 2009\footnote{The dataset has been collected in collaboration with the \textit{School of Informatics and Computing}, Indiana University at Bloomington, IN, USA.}. We explored a population of 99405 active users, extracting nearly 11 millions of annotations in the form of triples $\left\langle user, item, tag\right\rangle$ where an item can be an artist, an album or a specific song. From this raw dataset, we built the bipartite TRG, which has $1413657$ resource nodes and $285182$ different tags, from which we derived the FG.

\subsection{Last.fm dataset overview}\label{eval:overview}

\begin{figure}[tp]
\centering
\includegraphics[scale=.65, keepaspectratio]{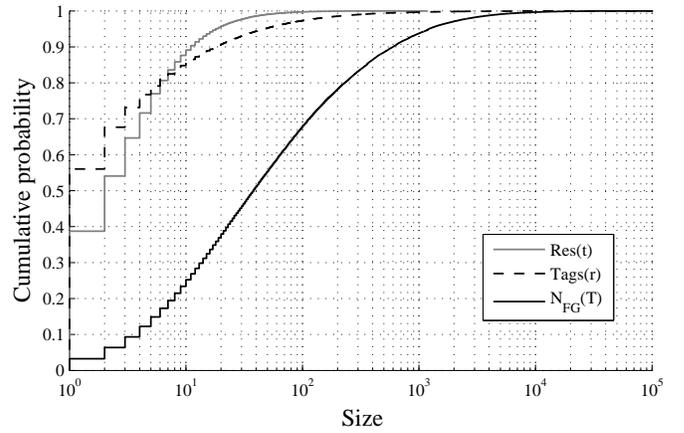}
\caption{Last.fm nodal degree CDF}
\label{fig:degree}
\end{figure}

We analyzed some structural properties of TRG and FG both. Some of the most relevant things to know are nodal degree distributions: in particular we extracted the distribution of the cardinalities of $Tags(r)$, $Res(t)$ and $N_{FG}(t)$ sets. Statistics of degrees (mean, standard deviation and max values, all rounded to integer) are shown in Table \ref{tab:stats} and cumulative degree distributions are depicted in Figure \ref{fig:degree}.

A strong core-periphery structure emerges in the TRG.
In particular, a huge portion of tags (about $55\%$) marks only $1$ resource and almost the $40\%$ of resources are labeled with just $1$ tag. Conversely, the dataset has a core of much more connected tags and resources; these correspond to the semantic top-level (or at least high-level) tags (e.g. ``rock'', ``pop'', ``seen live'') and to the most popular resources.

A similar scenario comes with the FG: the $80\%$ of tags has a not-null similarity with at most one or two hundred nodes, while the nodes belonging to the core of most popular tags are connected with several thousand nodes.

Given this setting, it is clear that the updates and lookup operations performed within the core structures of the network are the most ``problematic'' in terms of DHT operations. First, the number of tags marking a resource can be too high to avoid Approximation $A$. Second, given a very popular tag, the number of related tags and resources can be definitely huge; since, usually, overlay messages are sent on UDP packets, the limited payload force to send only a subset of tags and resources available during a search step. Therefore it is important that only the most relevant objects are returned; however, the definition of the DHT \textsc{get} primitive can easily be adapted with proper \textit{index side filtering} options in order to meet this requirement.
\begin{table}[tp]
\centering
	\begin{tabular}{|c|c|c|c|}
	  \hline
		\textbf{\textit{Degree}} & \textbf{\textit{Tags(r)}} & \textbf{\textit{Res(t)}} & \textbf{$N_{FG}(t)$}\\
		\hline
		$\mu$ & 5 & 26 & 316\\ 
		\hline
		$\sigma$ & 13 & 525 & 1569\\ 
		\hline
		$max$ & 1182 & 109717 & 120568\\
		\hline
	\end{tabular}
	\caption{Last.fm graph degree statistics}
	\label{tab:stats}
\end{table}

\subsection{Approximated graph simulation}\label{eval:experiments}

Given the Last.fm TRG and FG we simulate the evolution of such graphs with our approximated protocol in order to draw a comparison between the real dataset and the approximated one.

The simulation starts with a fully disconnected graph that includes all tags and resources from the Last.fm dataset. At each step, a resource $r$ and a tag $t$ are selected and a tagging operation is performed. The FG is updated according to Approximations $A$ and $B$. Resource $r$ is chosen with a probability proportional to its popularity in the dataset (i.e. $|Tags(r)|$ in the real TRG); tag $t$ is selected between all tags in $Tags(r)$ on a local popularity basis (i.e. with probability proportional to $u(t,r)$). Simulation ends when resources are labeled with all their related tags instances that appear in the real dataset. We executed the simulation for different values of connection parameter $k$.

\begin{figure}[tp]
\centering
\includegraphics[scale=.56, keepaspectratio]{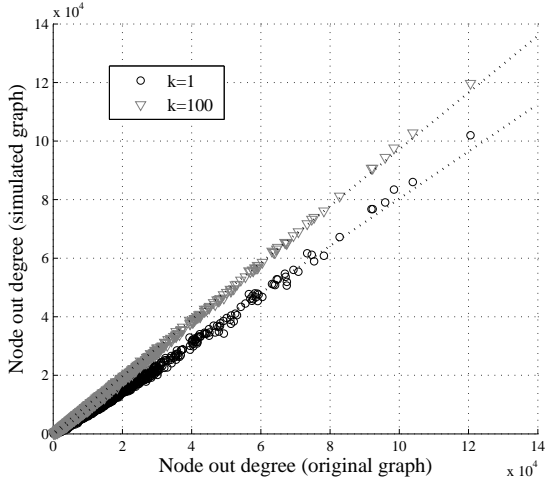}
\caption{Comparison between original and simulated FGs' nodal degree}
\label{fig:simuldegree}
\end{figure}

\begin{figure}[tp]
\centering
\includegraphics[scale=.57, keepaspectratio]{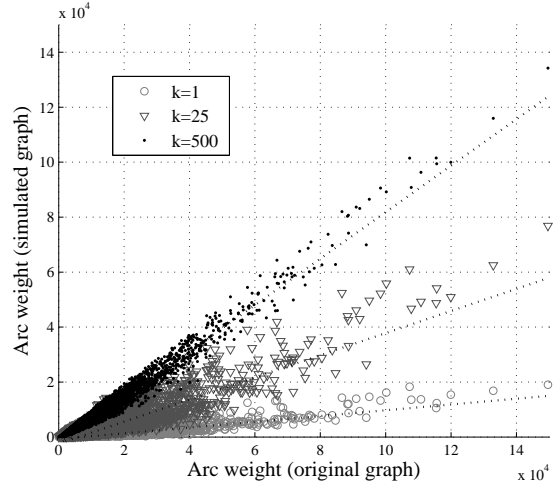}
\caption{Comparison between original and simulated FGs' arcs weights}
\label{fig:simulweight}
\end{figure}

We compare the original and the simulated FGs. We consider nodes out-degree and arcs weight (i.e. the tag-tag similarity values) values in original graph against corresponding values in simulated graph: the result are depicted in Figures \ref{fig:simuldegree} and \ref{fig:simulweight}.
We notice that, even with $k=1$, the points on the degree plot are aligned on a line whose slope is close to the diagonal; so we deduce that the variation of $k$ does not significantly affects the nodal degree. On the contrary, arcs' weight is significantly reduced for low values of $k$; to reduce the spread with the original values under a reasonable threshold, $k$ must be set to values that would make an efficient implementation on a DHT system unfeasible.

Nevertheless, what we are interested in is not to minimize the residues between theoretic and actual arcs' weight, but our aim is that some kind of \textit{proportions} are kept. First, we want that the arcs' weight ordering is maintained because the ranking of the $sim(t_1,t_2)$ weights directly influences the tags' ranking that is displayed during the search process. Second, we want that the proportion between the weight of every pair of arcs is not lost; if weight ordering is preserved for a pair of arcs but the ratio between their values significantly changes in the simulated network, then there is the risk of a \textit{flattening} effect on the tag similarity values, thus reducing the information provided to the user in the search step.

To give a quantitative measure of such advisable features, we compared, for each tag $t$ in the dataset, the set of its outgoing arcs $(t,t_i), t_i \in N_{FG}(t)$ with the same set taken from the approximated graph. The metrics we used for the arcs weights comparison are the Kendall's tau rank correlation coefficient ($K_\tau$) and the cosine similarity ($\theta$). $K_\tau$ evaluates the similarity between two ranks of a same set of objects on the basis of the number of inversions that have to be made to turn one ranking into the other; it ranges from $-1$ (when two rankings are the opposite) to $1$ (for equal rankings). $\theta$, which has the same range of $K_\tau$, takes in input two vectors of the same length and is equal to $1$ if these vectors are perfectly scaled (e.g. $\theta([1,2,3],[100,200,300]) = 1$).
\begin{figure*}[tp]
\centering
\includegraphics[scale=.57, keepaspectratio]{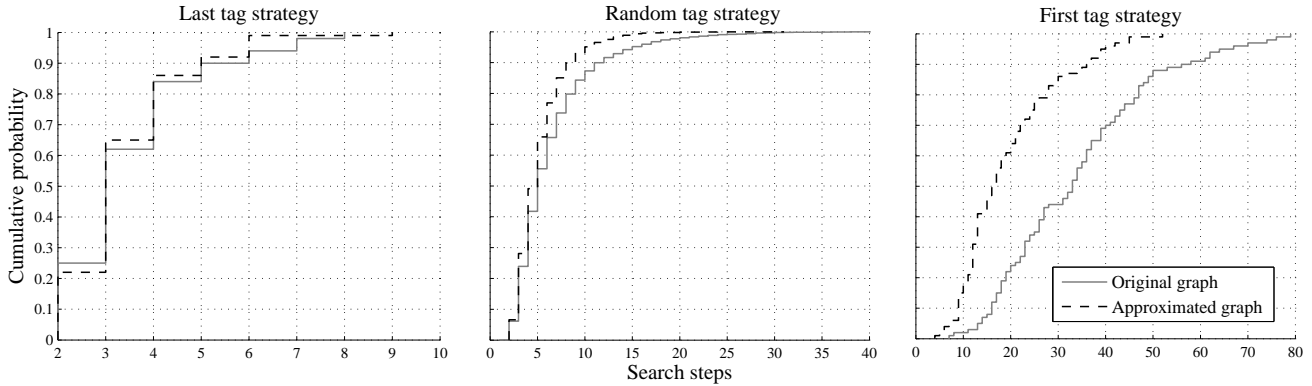}
\caption{Effect of approximation on tag navigation path length}
\label{fig:cdftot}
\end{figure*}

Furthermore, to give an estimation of how much information is lost with the approximation, we calculated the \textit{recall} value, that is the ratio between the number of arcs in the approximated graph and in the theoretical graph. Finally, we calculated the portion of arcs, among the set of those arcs that are not represented in the approximated graph, whose weight is $1$ in the theoretical model (we refer to this measure as $sim^1_{\%}$).
\begin{table}[bp]
\centering
\begin{tabular}{|c|c|c|c|c|c|}
\hline
\multicolumn{2}{|c|}{\textbf{\textit{k}}} & \textbf{Recall} & \textbf{$K_\tau$} & \textbf{$\theta$} & \textbf{$sim^1_{\%}$} \\
\hline
\multirow{2}{*}{$1$} & $\mu$ & $0.6103$ & $0.7636$ & $0.8152$ & $0.9214$ \\
\cline{2-6}
& $\sigma$ & $0.2798$ & $0.2728$ & $0.1978$ & $0.1044$ \\ 
\hline
\multirow{2}{*}{$5$} & $\mu$ & $0.7268$ & $0.7638$ & $0.8664$ & $0.9346$ \\
\cline{2-6}
& $\sigma$ & $0.2730$ & $0.2380$ & $0.1636$ & $0.0914$ \\ 
\hline
\multirow{2}{*}{$10$} & $\mu$ & $0.7841$ & $0.7985$ & $0.8971$ & $0.9432$ \\
\cline{2-6}
& $\sigma$ & $0.2686$ & $0.2138$ & $0.1424$ & $0.0850$ \\ 
\hline
\end{tabular}
	\caption{Comparison between approximated and theoretic Folksonomy Graph}
	\label{tab:comparison}
\end{table}

The mean values and the standard deviations of the cited measures, for some low value of $k$, are reported in Table \ref{tab:comparison}. The main results obtained are the following:
\renewcommand{\labelenumi}{\Alph{enumi}.}
\begin{enumerate}
	\item $K_\tau$ and $\theta$ values, measured on the set of tags which are common to the two models, are very high, independently on the value of $k$. This means that retrieved tags in the approximated model are well ordered and proportioned compared to the theoretical model.
	\item The value of $Recall$ reveals that, for very small values of $k$, up to the $40\%$ of arcs are not represented in the approximated model. Recall grows sub-linearly with $k$.
	\item The extremely high values of $sim^1_{\%}$ reveals that the weight of almost all these missing arcs is $1$, which is the minimum value in the similarity network. Further analyses showed that, for every $k$, the $99\%$ of the missing arcs has a weight $\leq 3$. So, the missing arcs are positioned in the very tail of the weight ranking.
\end{enumerate}

In a nutshell, even if correct proportions are kept, the number of arcs in the approximated FG can be considerably smaller respect to the original graph. Nevertheless, the arcs that are not mapped represent very weak similarities. In fact, the great majority of these arcs are simply \textit{noise} caused by the insertion of meaningless or singleton tags, which cover a high percentage of the overall tags but that are useless during the search phase. Therefore, the approximation adopted does not affect the quality of the FG, but rather reduces the noise on the mapped graphs and eases the load on the p2p layer.

\subsection{Faceted search's convergence}\label{eval:convergence}

Search convergence is important because the quickest the navigation converges, the lowest the number of overlay lookups needed to locate a resource is.

Convergence rapidity depends by which is the first tag selected. If it resides in the periphery of the FG, the search procedures will converge almost immediately, because the number of tags and the number of resources connected with it will be very probably quite small. Making a parallel with a taxonomic search structure, it is like the user had started his search from a node which is very close to a leaf of the tree structure, and so he/she had few levels left to explore.

On the contrary, the dual (and probably more frequent) behavior starts the search from more popular tags, those that resides into the core.
In order to show that convergence is quick also in this case we report further simulative results. We took the $100$ most popular tags and, starting from these, we simulated tag search procedures in order to estimate the average length of a search.

\begin{table}[bp]
	\centering
	\begin{tabular}{|c|c|c|c|c|}
		\hline
		\multicolumn{2}{|c|}{\textbf{\textit{Steps}}} & \textbf{Last} & \textbf{Rand} & \textbf{First} \\
		\hline
		\multirow{3}{*}{Original} & $\mu$ & 3.47 & 6.412 & 33.94 \\
		\cline{2-5}
		& $\sigma$ & 1.4175 & 4.4587 & 15.9942 \\
		\cline{2-5}
		& $\mu_{1/2}$ & 3 & 5 & 33 \\
		\hline
		\multirow{3}{*}{Simulated ($k=1$)} & $\mu$ & 3.38 & 5.2140 & 19.17 \\
		\cline{2-5}
		& $\sigma$ & 1.2373 & 2.6994 & 10.3065 \\
		\cline{2-5}
		& $\mu_{1/2}$ & 3 & 5 & 16 \\
		\hline
	\end{tabular}
	\caption{Search simulation statistics}
	\label{tab:searchStats}
\end{table}

Three types of search were performed; independently from the search strategy, we suppose that the size of the tag set shown to the user at each step, $T_i$, is upper bounded to the top $100$ tags retrieved from the DHT; larger sets of tags would be unsuitable for an effective user visualization. In the first search type (\textit{first tag strategy}) the tag selected at each step is the most similar with the current tag. Formally, given a search path $t_0, ..., t_i$, the next tag selected is a label $t_{i+1}$ such that $sim(t_i,t_{i+1}) \geq sim(t_i,\tau), \forall \tau \in T_i$. The second type (\textit{last tag strategy}) is the dual of the previous: the selected label is always the tag which is the least related with the current one among the $100$ tags displayed (i.e. tag $t_{i+1}$ such that $sim(t_i,t_{i+1}) \leq sim(t_i,\tau), \forall \tau \in T_i$). In the third search type (\textit{random tag strategy}), the next tag is selected uniformly random within $T_i$.

For each tag among the $100$ most popular we simulated the \textit{``first''} and \textit{``last''} search and $100$ \textit{random} searches, on both original and approximated Folksonomy Graph (for $k=1$), using the faceted search algorithm described in Section \ref{model:navigation}. The search procedure is stopped when $|T_i|$ reduces to $1$ or when $|R_i| \leq 10$. We choose $10$ as lower threshold for the number of displayed resourced because a set of $10$ objects is small enough to be displayed to the user without the need of further filtering.

Statistics on search paths length are shown in Table \ref{tab:searchStats}. From the experiments, it emerges that the path length is characterized by a high variance, for every search strategy, due to the high variability in the nodal degree of the FG.

With regard to searches performed in the original model, we observe that in the ``\textit{last}'' and ``\textit{random}'' strategy, the mean (and median) values are very small if compared to the size of the dataset; in particular, note that these values are $< ln(|T|)$. The ``\textit{first tag}'' strategy produces longer paths; however, a deeper result inspection revealed that they are originated by tag selection sequences which are very unlikely to be produced by a real user.

Roughly, the great majority of such sequences are those in which almost all tag selected are the most popular tags; since such tags are connected with huge sets of tags and resources, the size of the resource and tag sets decreases slowly at each search step. This is an expected behavior of the system, because if the user does not specialize the search terms it is clear that the navigation is maintained at a very coarse-grain level. Other slow-converging sequences are those in which many synonyms appear (e.g. ``electronica'', ``electronic'', ``electro''). Here, since semantically equivalent tags mark more or less the same set of resources, it is clear that navigation from one to another does not add any filtering information to the search procedure.

Such categories of search path occur because the meaning of the tag is not taken into account in simulations. But when the tag navigation is executed by a human user, and a semantic thread is followed in tag selection, the path leading to the objective could probably result shorter, more similar to the ``random tag'' selection case.

Comparing the simulative results obtained in the original Folksonomy Graph with those obtained for the approximated graph, the advantage on query convergence determined by approximation is clearly shown. Figure \ref{fig:cdftot}, which plots the cumulative density function of search path lengths for both models in the three strategies, together with statistics of Table \ref{tab:searchStats}, shows that the approximated approach shortens the navigation, thus quickening convergence. This effect, particularly evident in the ``\textit{first tag}'' strategy, is determined by the deletion of lightweight arcs from the graph. By wiping out the noisy connections the semantic distance between tags is increased, thus leading to a faster vocabulary specialization during the tag selection process.

As final consideration, remember that the simulated search ends when the set of resources reduces to an arbitrary threshold set to $10$, but if this value is raised, even slightly, path lengths could be considerably reduced.

\section{Conclusions and future works}\label{concl}

We presented an approximated approach for the maintenance of a folksonomy graph in order to make feasible a fully distributed implementation of a tagging system. In practice, we introduce a connection parameter $k$ which acts as an upper bound to the number of lookups executed on a DHT based system.

Simulative and analytic studies show that the approximated representation of the similarity graph does not upset the features of our theoretic Folksonomy Graph model, even for $k$ = 1. Besides, approximation can (1) largely mitigates overfitting phenomena, (2) significantly reduce the number of overlay operations for new tag insertion without degrading the user search experience. The information which gets lost in the approximated mapping is prevalently noise.

Furthermore, the property of search navigation acyclity and convergence, typical of taxonomical representations, is granted by our framework, even if a taxonomy is not explicitly built from the flat tag space. The efficiency of tag navigation convergence is shown by a simulative experiment on a large dataset from Last.fm. The approximated mapping reduces the average number of search steps, because the elimination of noisy similarity links between tags leads to a more effective filtering when new tags are selected during the navigation process. The overall approach leads to a better exploitation of the DHT layer. The low number of lookups needed during the insertion/search phases allows an efficient implementation of a tag-based, general-purpose indexing service over a structured p2p network.

Emulative and evolutionary analysis is planned in the next future. Indeed, the way in which our system reacts to particular evolutions deserves further investigation. In particular, we are planning to study if our approximated model hampers the emergence of new tagging trends: forthcoming tests will address the dynamics of different tag-resource patterns, and how the continuous activity of the community of users affects the adaptability of our p2p model.

\bibliographystyle{IEEEtran}

\end{document}